# Towards a disaster response system based on cognitive radio ad hoc networks

Noman Islam
Department of Computing and Technology
Indus University
Karachi, Pakistan
noman.islam@indus.edu.pk

Ghazala Shafi Shaikh
Department of Computing and Technology
Indus University
Karachi, Pakistan
ghazala.shafi@indus.edu.pk

*Abstract*— **This paper presents an approach towards disaster management based on cognitive radio ad hoc network. Despite the growing interests on cognitive radio ad hoc networks, not much work has been reported on using them for disaster management. This paper discusses opportunities for disaster management based on cognitive radio ad hoc networks. In this direction, the paper presents a novel technique for disaster detection based on Artificial Neural Network (ANN). The ANN is trained using backward propagation algorithm. An ANN-based spectrum sensing scheme is also presented. Finally, a service discovery scheme is presented for coordination during the time of disaster. The simulation of proposed approach has been performed in NS-2 simulator. The proposed approach shows very low false negative alarm rate using the proposed disaster detection system. The spectrum switching time of spectrum sensing scheme is also analyzed along with an analysis of latency of proposed service discovery scheme.**

## I. INTRODUCTION

Disasters have really affected the lives of human since existence. The Mediterranean earthquake in Egypt and Syria in 1201 A.D killed 1,100,000 people. Similarly, the Influenza Epidemic of 1917 killed 2,000,000 people [1]. There are several efforts being taken by countries (based on their capacity and wealth) to manage these disasters. Some of these efforts are manual while in other places, technology has been employed to address a disaster situation. Despite of the measures taken, no country has been able to manage disaster completely. The overall statistics indicate that the rate of disaster incidence is increasing and this trend will continue in future [1]. The possible reasons are climate changes and settlement of people in disaster prone areas.

The use of technology for disaster management has been explored in recent years. The "Golden 72 hours" principle states that people trapped in a disaster situation have a large chance to survive if they are rescued in initial 72 hours [2]. Hence, if appropriate communication infrastructure can be deployed and people are rescued earlier, a lot of lives can be saved.

The advancement of wireless technology during past few years has enabled their usage for solving various problems of lives. Specifically, the availability of ad hoc and peer-to-peer connectivity devices have led to various types of military and commercial applications such as ubiquitous computing, future combat systems and wild life monitoring. Despite of their wide scale applications, these networks have not been extensively used for disaster management.

Any Disaster management system involves [1]: i) **Mitigation**: reducing the occurrence of disaster ii) **Preparedness**: equipping the people to survive against disaster iii) **Response:** reducing the impact of disaster and iv) **Recovery**: taking lives back to normal situation. The focus of this paper is to use cognitive radio ad hoc network (CRAHN) for disaster response scenarios. In CRAHN, the nodes are equipped with cognitive radios that can switch their spectrum for communication. The nodes are organized in infrastructure-less fashion spontaneously. Because of their flexible nature, ease of deployment and the ability to use spectrum optimally, these networks are ideally suited for disaster management.

Before proceeding towards the discussion on proposed approach for disaster management, this paper first discusses the related area in this domain. The proposed work is then presented which is followed by implementation of proposed approach and results. The paper concludes with discussion on future work.

## II. LITERATURE REVIEW

The history of disaster management can be traced back to early days of mankind (Noah's Ark), when the Prophet Noah (Peace By Upon Him) was told to prepare for an approaching flood [3]. Manual efforts have been done since then to solve disaster management. Because of severity of the damages done due to disasters, the UN declared 1990 as the "International Decade for Natural Disaster Reduction (IDNDR)".

As far as work on disaster management using technology is concerned, only few solutions have been proposed in literature. In this direction, an early work based on ICT has been presented by Meissner et al. [4]. They identified various challenges and presented an integrated communication infrastructure based on information technology for disaster response and recovery. George et al. [5] have used ad hoc network for situation management by presenting a novel architecture called DistressNet. The proposal exploits distributed and collaborative sensing, topology aware routing and resource localization for disaster response.

The Workpad architecture [6] comprises front-end (first responders) andback-end layers (an integrated communication infrastructure) for disaster management. Lien et al. [2] have proposed a communication infrastructure based on Mobile Ad hoc Network (MANET) for disaster management. Different disaster management system has been proposed based on MANET in [7-9].

In recent proposals, CRAHN has been used for management of disasters. In this direction, a novel cluster-based scheme for post-disaster management has been presented in [10]. The solution employs CRAHN and 3-tier resource management model and claims robust communication in disaster scenario.

Khayami et al. [11] have presented a CRAHN solution for smart grid communication. The proposed approach discusses how a city wide network based on cognitive radios can be deployed for addressing communication problems during disaster. Sun et al. [12] have suggested the use of cognitive radio vehicles for disaster scenario as they can be quickly deployed.

However, it has been identified that the work on disaster management is at infancy and no effective disaster management solution is currently available.

III. PROPOSED DISASTER MANAGEMENT SYSTEM

This work proposes a novel disaster management system based on CRAHN. The next few paragraphs discuss the proposed approach.

*A. Disaster detection*

Fig. 1 shows the block diagram representing proposed disaster detection system.

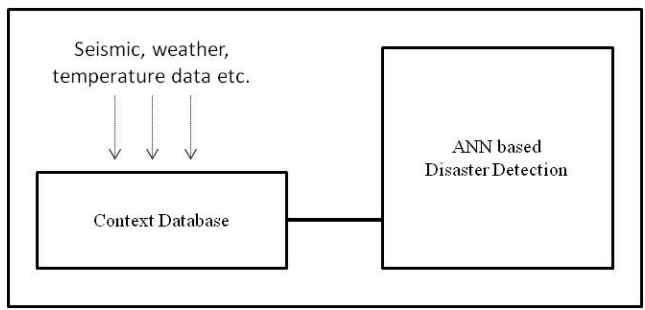

Fig. 1. Disaster Detection System

The area prone to disaster is equipped with sensors (smoke detector, seismometers and radar systems) to detect any disaster situation. The sensor data is passed on to a context manager that maintains status at different sites. The Artificial Neural Network (ANN) based disaster detection system uses context manager's data to decide whether a disaster is anticipated or not. Fig. 2 shows how ANN is used for this purpose. ANN is a three layer neural network that is first trained using feedback propagation algorithm in offline mode. The ANN classifies various contextual parameters such as temperature, weather and seismic data as disasters of various type in real-time.

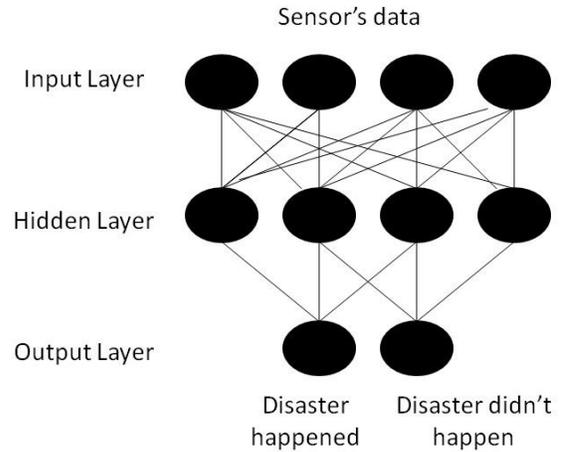

Figure 2: Using ANN to detect disaster

*B. Spectrum sensing*

A disaster naturally results in destroying the whole communication infrastructure such as: base stations are crashed, base station to Mobile Switching Center (MSC) connectivity is broken, power generators are exhausted, cell phones ran out of battery and network suffers from congestion [2]. In such systems, there is less than 1 percent chance of survival of existing communication infrastructure, and hence there must be improvised and pervasive communication used in such situation [2, 5]. The communication protocols should operate in decentralized fashion without availability of infrastructure (such as DHCP servers, directory services and security servers) [13]. As far as spectrum is concerned, The Industrial, Scientific and Medicine (ISM) band can be used for communication. However, such capacity can be exhausted due to the demands of disaster management system. Cognitive radio communication should be used in such situations.

Hence, this work proposes that nodes operating in the region will form a CRAHN for communication. The use of CRAHN enables the nodes to operate in licensed spectrum of other users (primary user), thus solving the problem of scarcity of the available spectrum.

Fig. 3 shows the system running at every node during the time of disaster. The cognitive radios installed in the network will detect spectrum holes. In addition, the system logs primary user spectrum usage which is then used by an ANN based spectrum manager to select the best spectrum hole. Every node also maintains a situation database characterizing the whole situation of disaster site.

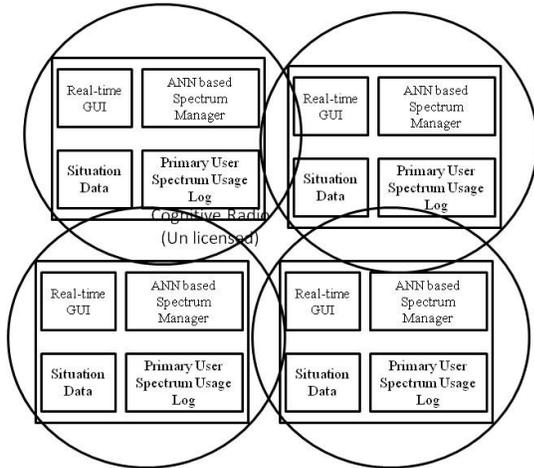

Figure 3: Cognitive Radio based communication during disaster

### C. Response Initiation

Upon detection of a disaster, the response phase is activated. An appropriate call-tree is used to inform the right person in case of a disaster, as normally done for disaster recovery. Listing 1 shows the pseudo code. To inform the corresponding personnel, a gateway node is required through which communication with external world can be established. The proposed approach recommends a discovery protocol for this purpose based on Islam and Shaikh [14] as discussed in detail in next section. A request for discovery of gateway service is floated on the network that is floated on the network until a node providing gateway service is found. This gateway node is then used for communicating about disaster to responsible personnel. First, the personnel at level 1 is informed and if no response is received for 10 second, the next level node is tried on. This process continues on until a personnel is found.

```
Listing 1: Pseudo code for Disaster Response Initiation
ContextManager cm = Runtime.getContextManager()
ANNDetector d = new ANNDetector(cm)
If(d.diasterExpected()) {
    Detect gateway node
    n = n + 1
    While(no Response from Personnel) {
        If all personnel contacted with no response Goto Alarm
        Call personnel at call tree level n
        wait(10 * 1000)
        n = n + 1
    }
    Level n officer declares disaster
    Alarm: Disaster alarm generated
    Disaster recovery initiated
}
```

### D. Managing disaster

Once a disaster response is initiated, various types of rescue services are launched. These services are available in the network and can be discovered/ invoked by injured persons to inform about themselves to rescue persons. Similarly, the service discovery algorithm can be utilized by rescue workers to know current location of another rescue worker or amount of explosive gases in the air [4]. The services information is used by nodes to extract situation data which is then saved in situation database.

### E. Service discovery algorithm

In the proposed approach, the node proactively broadcast services to neighboring nodes that are cached by receiving nodes. So, most of the services can be served from the rescue workers local cache. When a node needs a service that is not available locally, a service discovery request is floated on the network using the AODV routing protocol as discussed in Islam and Shaikh [14]. The request is propagated by intermediate nodes until requested services are found out. The response is then generated which travels back through the route used for request initiation. The advantage is that a route towards the service node is available along with the discovery process.

### F. Real-time data delivery

Since time is a critical factor to save lives, any disaster situation demands near real-time exchange of data among the nodes [5]. Currently, most of the disaster management system use manual tables and charts that are updated via verbal messages sent through radios. However, this approach is not feasible as such a situation requires real-time as well as on-demand situation awareness. In addition, support for real-time decision making is also required.

To address this problem, this work proposes that the rescue personnel have real-time application installed on their laptops (or computing gadget) and they can view the current situation. The GUI will comprise information as follows a) location b) situation whose values are green/yellow/red, c) timestamp d) a short message and e) a detailed message. If a help is required by rescue worker, it can post a situation message with current location, situation as "red", short message as "help" and detailed message describing the help required. The rescue workers can use the real-time display to make decisions, can take appropriate actions etc.

The nodes operating in the disaster region should be interoperable and the semantic differences should be resolved autonomously. To resolve these interoperability issues, a common XML format is used to maintain and disseminate situation data.

## IV. IMPLEMENTATION DETAILS & RESULTS

The proposed approach has been implemented in NS-2 [15]. The simulation is performed in an area of 1000 × 1000

m$^2$. The number of PU is set to 5, while SU is set to 50. The spectrum is set to 2.4 GHz. The mobility model used is random way point model.

Fig. 4 shows the false alarm rate for disaster detection when different number of input nodes *n* in ANN. The data from various sensors at site is aggregated in *n* groups and passed as input to ANN. The number of hidden nodes is set to 5.

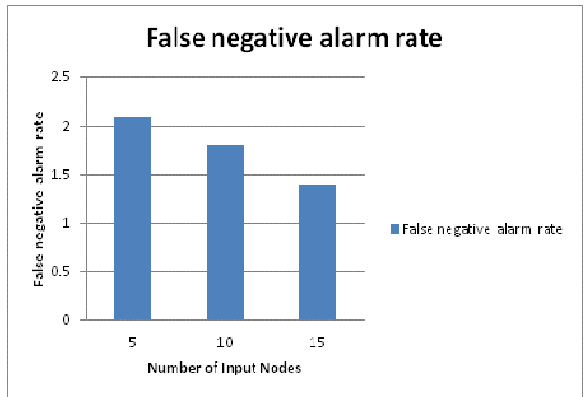

Figure 4: False negative alarm rate under different input nodes

Fig. 5 shows the spectrum switching time of various secondary users, when the primary user uses its spectrum after every 5 seconds. The average spectrum switching time is 1.32 sec.

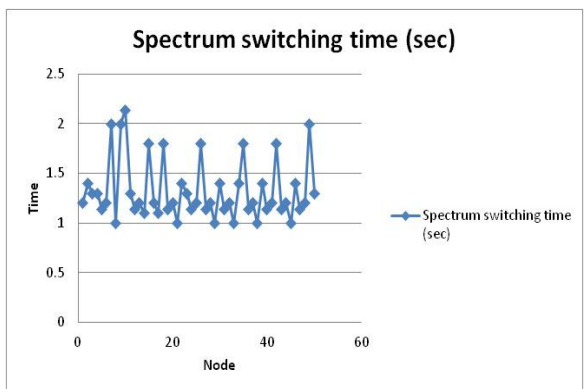

Figure 5: Spectrum switching time of secondary users

Fig. 6 shows the latency of various nodes when the service discovery scheme has been used. The number of services is set to 10. The average latency to discover a service on the network is 8.74 sec.

## V. CONCLUSION

In this paper, a disaster management framework has been proposed for cognitive radio ad hoc network. An ANN based disaster detection scheme, ANN based spectrum sensing approach and a service discovery scheme have been presented. The performance of proposed approach is analyzed using NS-2. The future work comprises testing of the proposed framework in some real-world scenario and testing for large number of nodes.

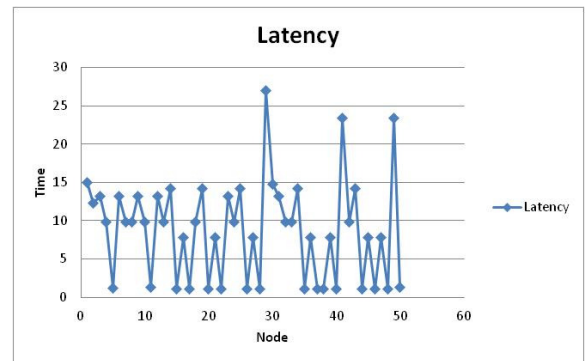

Figure 6: Avg. Latency of various nodes during service discovery